# Temperature dependence of the Casimir force between plane-layered structures: an approach based on the principle of argument

M.V. Davidovich

Saratov National Research State University named after N.G. Chernyshevsky, 410012, Saratov, Astrakhan 83, Russia

E-mail: davidovichmv@info.sgu.ru

Based on the complex analysis theorem “the principle of argument”, a temperature-dependent formula for the Casimir-Lifshitz dispersion force of the interaction of two arbitrary plane-layered structures separated by either a vacuum gap of size d or a layer of size d consisting of an arbitrary dissipative dielectric with dispersion is derived. The formula consists of the temperature-dependent Lifshitz formula plus the temperature-dependent contour integral. The results obtained are a generalization of the results of van Kampen and Schram for dispersion forces at zero temperature, which contains only a contour integral. It is shown that for a finite temperature in the case of ideally conducting structures (with infinite dielectric permittivity) with a vacuum gap, the results correspond to the results of the work of Mehra and Schwinger both at a distance tending to zero and at a distance tending to infinity. In the first case, the results coincide with Casimir 's result $F_C(d)=\hbar c\pi^2/(240d^4)$, and in the second case with the well–known formula $F_T(d)=k_B T/(4\pi d^3)$.

In [1], Casimir obtained the law of attraction of two ideally conducting plates with a specific (per unit area) force $F_C=\hbar c\pi^2/(240d^4)$. He considered the ideal problem of the interaction of two ideally conducting half-spaces through a vacuum gap of size *d*, summed up the infinite number of quanta $\hbar\omega_n/2$ of zero-point oscillations of the quantum vacuum in a resonator of large sizes *D* and *L* Fig. 1, and then, from the difference of two infinite quantities: the free energy density in the gap and the free energy density of unlimited space, assuming transparency of the structure at high frequencies, using the Euler-Maclaurin formula, he isolated the finite energy density $E_C = E/L^2 = -\hbar c\pi^2/(720d^3)$ and the attractive force $F_C = -\partial_d E_C = \hbar c\pi^2/(240d^4)$. In the model of ideal conductivity, instead of infinite half-spaces, it is quite possible to consider perfectly conductive sheets (screens) of zero thickness. Then the resonator considered by Casimir has a space between them of size *d*, Fig. 1. However, it is necessary to take into account the energy and pressure outside the structures (in the remaining part of the resonator or in the free space at $D \to \infty$). Casimir took this into account by subtracting the energy inside the gap from the energy outside the gap, taking it into the subtracted result. Obviously, this is equivalent. Applying the Euler-Maclaurin formula, Casimir actually tacitly put $d \to \infty$, i.e. took $k_m d >> 1$, where the distance *d* is significantly greater than the minimum wavelength $\lambda_{\min} << d$ below which the body can be considered transparent. This means that although Casimir used resonant frequencies of a perfectly conducting resonator, he actually obtained the result for real metals at infinitely large *d* by taking a large distance in the formula for $\delta E$. The model of a perfectly conducting body implies that for all wavelengths, including infinitesimal ones, it reflects perfectly. An ideal conductor can be modeled as a metal, the plasma frequency of which tends to infinity. Although Casimir used the pruning factor $f(k/k_m)$, he actually proved that the law $F_C = \hbar c\pi^2/(240d^4)$ is valid for all distances in the model of ideally conducting bodies if $k_m \to \infty$. This law also follows from the results of other authors who used other approaches to derive the Casimir formula.

In [2], Mehra considered a resonator in Fig. 1, region 3, separated by a perfectly conductive partition 1 and filled with equilibrium thermal radiation. He used the methods of statistical mechanics, determining the free energy of the resonator $E = -\beta^{-1}\ln(Z)$ through the statistical sum $Z = Sp(\exp(\beta H))$, the Hamiltonian $H = \hbar\sum_n \omega_n (N_n + 1/2)$ and the inverse dimensionless temperature $\beta = (k_B T)^{-1}$ in the form

$$E = \sum_n \left[\hbar\omega_n/2 + \beta^{-1}\ln(1 - \exp(-\hbar\omega_n\beta))\right]. \quad (1)$$

In the case $T \to 0$ the Casimir's result from (1) follows immediately: $E = \hbar\sum_n \omega_n/2$. In the case $T \to \infty$ from (1), after decomposition of the logarithm, it follows in the form

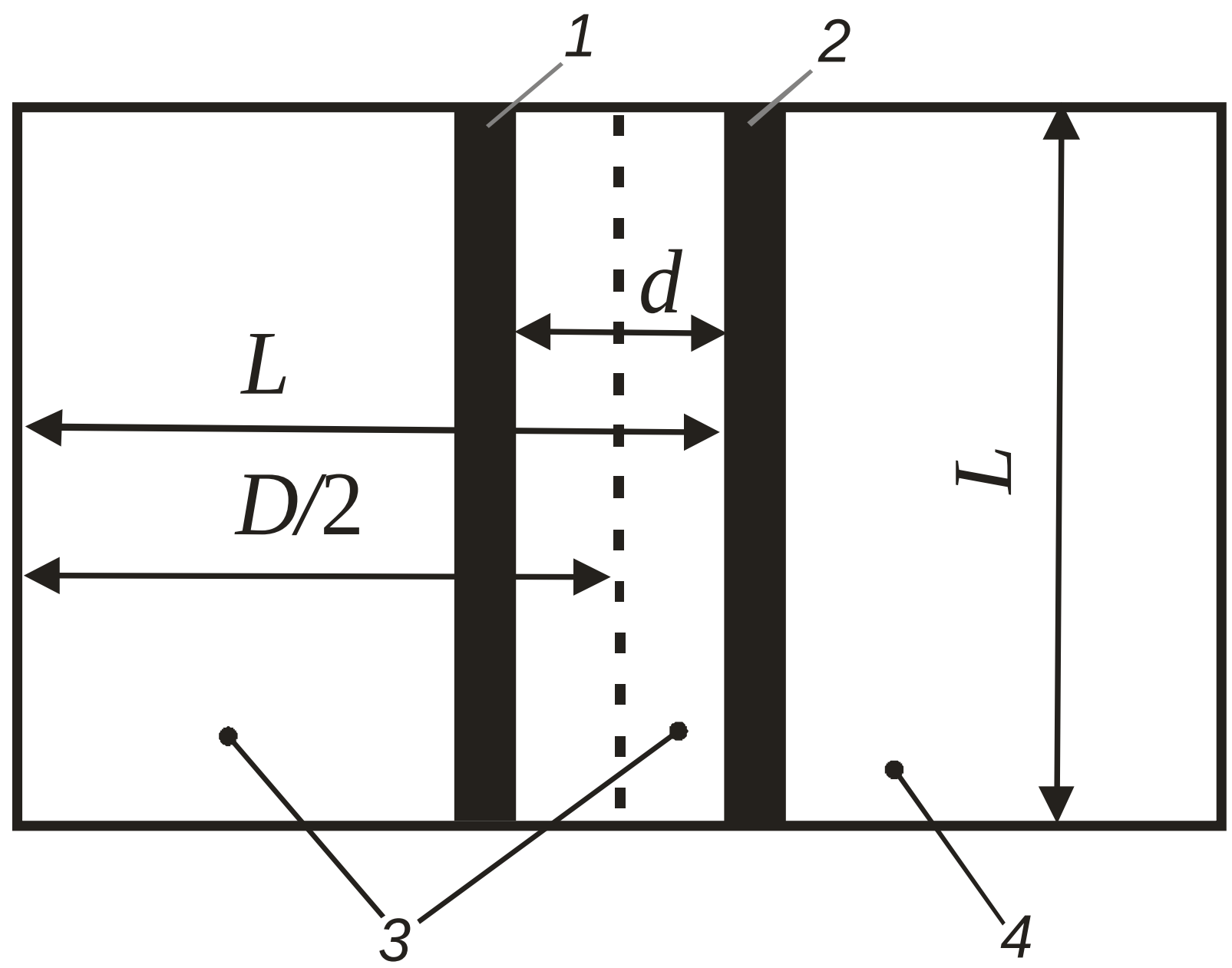

Fig. 1 Rectangular *D*x*L*x*L* resonator in two parts 3 and 5 with two ideally conducting structures 1 and 2 for the Casimir problem [1] and the *L*x*L*x*L* resonator for the problem [2]

$$E = \sum_n \left[ \hbar\omega_n / 2 - \left(\beta^{-1} / m\right) \sum_{m=1}^{\infty} \exp\left(-m\hbar\omega_n \beta\right) \right].$$

It is assumed here that $\hbar\omega_n < k_B T$ for all frequencies. Since at $T \to \infty$ in this formula the argument of the exponent changes by an infinitesimal amount, the series can be replaced by an integral: $dm = dx\left(k_B T / \hbar\omega_n\right)$, and then

$$\sum_{m=1}^{\infty} \frac{1}{m} \exp\left(-\frac{m\hbar\omega_n}{k_B T}\right) = \int_{\frac{\hbar\omega_n}{k_B T}}^{\infty} \frac{\exp(-x)}{x} dx = -\mathrm{Ei}\left(-\frac{\hbar\omega_n}{k_B T}\right) =$$

$$= -C - \ln\left(\frac{\hbar\omega_n}{k_B T}\right) - \sum_{m=1}^{\infty} \frac{(-1)^m}{m!m} \left(\frac{\hbar\omega_n}{k_B T}\right)^m .$$

All frequencies should be summed to the left of the partition

$$\omega_n^- = c\sqrt{\left(\frac{n_x \pi}{L}\right)^2 + \left(\frac{n_y \pi}{L}\right)^2 + \left(\frac{n_z \pi}{L-d}\right)^2}$$

and to the right of the **partition**

$$\omega_n^+ = c\sqrt{\left(\frac{n_x\pi}{L}\right)^2 + \left(\frac{n_y\pi}{L}\right)^2 + \left(\frac{n_z\pi}{d}\right)^2} ,$$

for all types of oscillations ($E_n$ and $H_n$), and to apply the assumption used above about the smallness of the exponential argument, an infinite frequency spectrum should be limited at a large (but fixed) temperature.: $\hbar\omega_n < k_B T$. Changing the position $\delta d$ of the partition changes the energy $\delta E = -(\partial_d E(d) - \partial_d E(L-d))\delta d$, which leads to a specific force $F = \partial_d E / L^2$. Next, Mehra, like Casimir, moved on to an open resonator at $L \to \infty$ and found a force at low temperature $d/\delta_T << 1$ in the form of

$$F(d,\delta_T) \approx \frac{\hbar}{2L^2}\sum_n \partial_d \omega_n = F_C(d)\left[1 - \frac{240 d}{\pi\delta_T}\exp\left(-\frac{\pi\delta_T}{d}\right) + \frac{16}{3}\left(\frac{d}{\delta_T}\right)^4\right] \tag{2}$$

and at high temperature $d/\delta_T >> 1$ in the form of

$$F(d,\delta_T) \approx \frac{k_B T}{L^2}\partial_d \sum_n \ln\left(1 - \exp\left(-\frac{\hbar\omega_n}{k_B T}\right)\right) = \\ = \frac{k_B T \varsigma(3)}{4\pi d^3}\left[1 + 2\left(1 + \frac{4\pi d}{\delta_T} + \frac{1}{2}\left(\frac{4\pi d}{\delta_T}\right)^2\right)\right]\exp\left(-\frac{4\pi d}{\delta_T}\right). \tag{3}$$

The thermal length $\delta_T = \hbar c/(k_B T)$ is entered here. To highlight Casimir's result, Mehra applied the same approach as in [1], i.e. assumed a limited opacity spectrum of screens, introduced a cropping factor *f* with a maximum wavenumber $k_m$, and used the Euler-Maclaurin formula. However, if we will strive $k_m \to \infty$ in cropping factor, then result (2) is valid for ideal screens at $d \to 0$. At a fixed temperature, the limit must be met $F(d,\delta_T) \to F_C(d) = \hbar c \pi^2/(240 d^4) \to \infty$, $d \to 0$. At a fixed temperature from (3) at $d \to \infty$ we have $F(d,\delta_T) \to k_B T \varsigma(3)/(4\pi d^3) \to 0$. This law of decreasing force is slower than in $F_C$, i.e. the thermal field increases the force. For small d, there is also a slight increase in strength compared to $F_C$, since the exponent in (2) is very small. Formulas (2) and (3) were confirmed by Schwinger et al. [3]. Schwinger used the method of tensor Green functions (GFs) and the general variational principle of electrodynamics. Prior to Mehra, the results (2) and (3) were obtained in 1962 by Sauer in an unpublished dissertation (see [3,4]). These results were also confirmed in [4] based on the GFs obtained by the imaging method.

In [5], van Kampen et al. and Schram [6] used the complex analysis theorem "principle of argument" to derive the Casimir force (see also [4,5,7,8]). In [5], the formula was obtained without taking into account the delay, and in [6] – with taking into account the delay. The formula with a delay has the form

$$F(d) = \frac{\hbar}{2\pi^2 c^3} \int_0^\infty \xi^3 d\xi \int_1^\infty p^2 dp \left[ \frac{1}{f_e(\xi, p, d)} + \frac{1}{f_h(\xi, p, d)} \right]. \tag{4}$$

In fact, even before [6], the formula was obtained in [9]. Gerlach obtained the formula by considering the interaction of plasmon-polaritons (PP) along the surfaces of dissipative structures. In fact, he derived formula (4) using the fact that the force is caused by the interaction of charges and currents induced on the surfaces during the movement of the PP. In (4), the integral over $p$ defines integration over the transverse wavenumbers, and integration over $\xi = i\omega$ corresponds to integration over the imaginary frequency axis. For the introduced functions we can obtain [6-10] $f_\mu(\xi, p, d) = r_{1\mu}^{-1} r_{2\mu}^{-1} \exp(2\xi p d) - 1$, $\mu = e, h$. They indicate the coefficients of reflections from the gap side from two structures on the left and right for E-waves ($\mu = e$) and for H-waves ($\mu = h$). For symmetrical structures $r_{1\mu}^{-1} r_{2\mu}^{-1} = r_\mu^{-2}$. The entire variety of structures under consideration is described only by these reflection coefficients. Next, we will also use an imaginary wave number $k = ik_0 = i\omega/c$. In the case of a vacuum gap there is a transverse wave number $\kappa = \sqrt{k_x^2 + k_y^2} = k\sqrt{p^2 - 1}$, therefore $k_z = \sqrt{k_0^2 - \kappa^2} = -ikp$. However, the formula is also applicable when filling the gap with an arbitrary dielectric with a dielectric permittivity (DP) $\tilde{\varepsilon}(\omega)$. Then $\kappa = k\sqrt{\tilde{\varepsilon}(k)}\sqrt{p^2 - 1}$ and $k_z = -ikp\sqrt{\tilde{\varepsilon}(k)} = -iK$. Formula (4) coincides with the formula given in Lifshitz's work as early as 1955 in the reports of the USSR Academy of Sciences and in [10], published in a translated version in 1956. The case of two dielectric spaces separated by a vacuum gap is considered there. In the Lifshitz configuration $r_e = (s - \varepsilon p)/(s + \varepsilon p)$, $r_h = (s - p)/(s + p)$, $s = \sqrt{p^2 - 1 + \varepsilon}$, $\xi = i\omega$. We use dependence $\exp(i\omega t)$. To switch to dependence in [10], a substitution $i \to -i$ should be made. Lifshitz's formula has caused a boom in theoretical and experimental studies of dispersion forces or van der Waals–Casimir Lifschitz forces. This formula for $r_\mu^2 = 1$ (ideal screens) implies the result of Casimir [1–4,7–10], and for arbitrary d. Formula (4) for the configuration of work [10] was proved in [3,11]. Levin and Rytov [11] used the theory of equilibrium thermal fluctuations in electrodynamics, the excitation of a structure by dipoles, the finding of field correlations and their introduction into the Maxwell stress tensor. They applied the solution of the Sommerfeld problem on the excitation of dielectric half-spaces by electric and magnetic dipoles in the gap between them and obtained the Lifshitz formula. Schwinger and his co-authors used the tensor functions of the GFs structure and the general variational principle of electrodynamics for a plane-layered structure. The works of [5,6,9] actually generalize the result of [10] to arbitrary plane-layered structures, while Gerlach used the symmetry properties of the functions $f_\mu(-\xi,-p,d) = f_\mu(\xi,p,d)$, obtained by him for a multilayer structure. The equations $f_\mu(-\xi,-p,d) = f_\mu(\xi,p,d) = 0$ are nothing more than the dispersion equations of PP in the structure. This means that $f_\mu(-\omega,-p,d) = f^*(\omega,p,d) = 0$, even then, the phase shift persists when the direction of motion and frequency change. Similar symmetry conditions hold for DP and conductivity, for example $\varepsilon_{\alpha\beta}(-\omega,-\mathbf{k}) = \varepsilon^*_{\varepsilon\beta}(\omega,\mathbf{k})$, where the real part of DP is even and the imaginary part is an odd function [12]. When moving to an imaginary frequency, DP becomes a real function with symmetry $\varepsilon_{\alpha\beta}(-\xi,-\mathbf{k}) = \varepsilon_{\alpha\beta}(\xi,\mathbf{k})$. In the case of a vacuum gap, (4) includes a function $\exp(2\xi pd) = \exp(-2i\omega d/c)$ that means a phase shift during double passage of the gap in time $\tau = 2d/c$. Therefore, the Lifshitz formula (4) has a simple physical meaning. Decomposing the fractions in (4) into the Neumann series [13], we see that the interaction is caused by multiple radiations and absorption of reflected waves by bodies. Zero reflection coefficients result in zero strength. If $r_\mu = -1$ (ideal electric screens), then the pressure of the quantum vacuum in the gap is maximum. It is caused by standing waves. However, outside the bodies, they are affected by external pressure, which is greater than the internal pressure and does not depend on *d*. It is caused by waves running towards the body from infinity and reflected by it. When the plates slide apart $d \to \infty$, the external and internal pressures are balanced, and the force is reduced to zero. The same is true for two ideal magnetic screens $r_\mu = 1$. However, in the case of an ideal one magnetic and one electric shield, the force changes sign, becoming repulsive: $F = -7\pi^2\hbar c/(1920 d^4)$ [14]. The sign of the force is determined by the phase relations. The transmission of a pulse to a body from the side of the vacuum gap and from the outside depends on both absorption and reflection, i.e. on the modulus and phase of the reflection coefficient. If at a certain frequency $r_{1\mu} = 0$, the neighborhood of that frequency does not contribute to the force. For zero oscillations, this means that their boundary conditions do not change both inside and outside the body, i.e. the body is transparent at a given frequency for waves of both directions formed by quantization of the electromagnetic field by the imposition of periodic boundary conditions. In the case of thermal fluctuations, the body at a frequency for which $r_{1\mu}=0$ is completely absorbs photons, both from the gap side and from the outside. But according to the Kirchhoff principle, it emits them as a completely black body (at a specified frequency), i.e. it does not change the spectral energy density at a given frequency. The second body at the specified frequency does not seem to feel the influence of the first. In general, this effect is determined by the phases and modules of the reflection coefficients. In a dissipative substrate, the wave remains attenuated in the direction of motion during the transformation $f_\mu(-k,-p,d)=f(k,p,d)$. Formula (4) can be reduced to an integral with infinite limits

$$F(d)=\frac{\hbar}{8\pi^2c^3}\int_{-\infty}^{\infty}\xi^3\,\mathrm{sgn}(\xi)d\xi\int_{|p|>1}p^2dp\left[\frac{1}{f_e(\xi,p,d)}+\frac{1}{f_h(\xi,p,d)}\right],$$

by representing it as an integral over both positive and negative frequencies $\xi$. Considering it only for positive *p*, we have

$$F(d)=\frac{\hbar}{4\pi^2c^3}\int_{-\infty}^{\infty}\xi^3\,\mathrm{sgn}(\xi)d\xi\int_{p>1}p^2dp\left[\frac{1}{f_e(|\xi|,p,d)}+\frac{1}{f_h(|\xi|,p,d)}\right].$$

In this formula, we can take $\xi^3\,\mathrm{sgn}(\xi)=\xi^2|\xi|$, i.e. the integral function in (4) is considered as an even function of $\xi$.

Since for metal plates, the DP $\varepsilon(\omega)=\varepsilon_L(\omega)-\omega_p^2/(\omega^2-i\omega\omega_c)$ and surface conductivity $\sigma(\omega)=\sqrt{\varepsilon_0\varepsilon(\omega)/\mu_0}$ become infinite only in the limit $\omega\to 0$, real metals interact with the Casimir force at large distances, and at small ones the force must be finite. The specific conductivity included in Maxwell's equations is defined in terms of DP as $\gamma(\omega)=i\omega\varepsilon_0(\varepsilon(\omega)-1)$. Here we use time dependence $\exp(i\omega t)$. The value $\varepsilon_L(\omega)$ is determined by the Lorentz dispersion. It is significant in the UV range, where the Drude term is small. The main contribution to the force is made by frequencies with a spectral range $\omega\sim c/d$, therefore, taking into account the Lorentz terms is important for distances up to 300 nm. For lower frequencies, we can assume that $\varepsilon_L(\omega)\sim 10$ and is independent of frequency. In this case, the interaction of metals for distances $0.3 < d < 10$ μm can be described by a plasma model without taking into account dissipation. The Drude model should be used for distances $d > c/\omega_c$, and dissipation is important.

So, formula (4) for zero temperature was proved in [3,5,6,9,11] and reliably established experimentally [7]. Casimir's result follows from it for $r_\mu \to \mp 1$ and, importantly, for any *d*. For ideally conducting screens (a Fabry-Perrault resonator) with a finite temperature, this formula was obtained in [2–4], and both at $T \to 0$ and finite *d*, and at $d \to 0$ and finite temperature, the Casimir result follows from (2). Thermal photons do not affect the force at small distances, since they are emitted by the body into the far wave band [11].

*The purpose of this work* is to obtain a formula for the dispersion interaction of two arbitrary plane–layered structures based on the principle of the finite temperature argument. To do this, we determine the free energy of the resonator in Fig. 1 using the energy of a quantum oscillator in a thermal field $\Theta(\omega,T) = \hbar\omega \coth(\omega_n \hbar/(2k_B T))/2$ [11]. Then

$$E(d) = \frac{\hbar}{2}\sum_n \left( \omega_{ne}(d) \coth\left( \frac{\hbar\omega_{ne}(d)}{2k_B T} \right) + \omega_{nh}(d) \coth\left( \frac{\hbar\omega_{ne}(d)}{2k_B T} \right) \right).$$

The summation proceeds in electric and magnetic modes, $n = (n_x, n_y, n_z)$. In the case of temperature, a characteristic parameter arises: $\delta_T = \hbar c/(k_B T)$ – the thermal length. Then in the case $T \to 0$ we have the Casimir result. In the case $T \to \infty$

$$\coth\left( \frac{\hbar\omega_{ne,h}(d)}{2k_B T} \right) = \frac{2k_B T}{\hbar\omega_{ne,h}(d)} + \frac{1}{3}\left( \frac{\hbar\omega_{ne,h}(d)}{2k_B T} \right) + \ldots \approx \frac{2k_B T}{\hbar\omega_{ne,h}(d)},$$

therefore, when summing up to a first approximation, the dependence on *d* and on $\hbar$ disappears. Since the force is obtained by differentiation in *d*, this term is not significant, and decompositions should be obtained at a large but finite temperature. Therefore, the free energy and its derivative with respect to *d* should be calculated more carefully. In the case $T \to 0$

$$\coth\left( \frac{\hbar\omega_{ne,h}(d)}{2k_B T} \right) = 1 + 2\exp\left( -\frac{\hbar\omega_{ne,h}(d)}{k_B T} \right),$$

therefore, in the first approximation, the dependence on temperature disappears. When passing to an open resonator $dn_x = L dk_x/\pi$, $dn_y = L dk_y/\pi$, $n_z = n$, and we have

$$E(d) = \frac{\hbar L^2}{4\pi}\int_0^\infty \kappa d\kappa \left[ \sum_n \coth\left( \frac{\hbar\omega_{ne}(\kappa)}{2k_B T} \right)\omega_{ne}(\kappa) + \sum_n \coth\left( \frac{\hbar\omega_{nh}(\kappa)}{2k_B T} \right)\omega_{nh}(\kappa) \right]. \quad (5)$$

In this case, (5) summarizes the energy over all the dispersion branches of the electric and magnetic PP. There are slow surface and fast leakage PP. In the absence of dissipation for ideally conductive screens, the DEs passes into characteristic equations. And in this case $k_z^2 = (n\pi/d)^2 = k_0^2 - \kappa^2$, $\omega_{n(e,h)} = c\sqrt{\kappa^2 + (n\pi/d)^2}$, while the modes become radiated volumetric with a real continuous frequency spectrum. In a shielded resonator, the spectrum is discrete.

Let's assume that instead of plates in Fig. 1 there are two infinitely thin sheets with surface conductivity, for example, two sheets of graphene. Graphene conductivity is approximately described by the Drude model [15], and it is performed for it $\sigma(-\omega) = \sigma^*(\omega)$. For the reflection coefficients from the side of the vacuum gap $r_{e,h}(\omega) = -\sigma(\omega)/(\sigma(\omega) + 2y_{e,h}(\omega)) = r_{e,h}^*(\omega)$, where the dimensionless wave conductivities in vacuum of E-waves $y_e(\omega) = k_0/k_z$ and H-waves $y_h(\omega) = k_z/k_0$ are indicated . Therefore $r_{e,h}(-\omega) = r_{e,h}^*(\omega)$, if $k_z(-\omega) = k_z^*(\omega)$. This means that the forward wave $\exp(i\omega t - ik_z(\omega)z)$ remains the same when the frequency sign changes, while attenuation always occurs in the direction of propagation. Changing the sign of the frequency means changing the direction of time, i.e. moving towards –*z*. In particular, if the wave is surface $k_z(\omega) = -iK$, $k_z(-\omega) = iK$, i.e. there is an increase in the *z* direction (the wave is opposite). In the case of a dielectric or metal plate

$$r_e(k,p) = -\frac{|p|\varepsilon(k) - \sqrt{\varepsilon(k) + p^2 - 1}}{|p|\varepsilon(k) + \sqrt{\varepsilon(k) + p^2 - 1}},$$

$$r_h(k,p) = -\frac{\sqrt{\varepsilon(k) + p^2 - 1} - |p|}{\sqrt{\varepsilon(k) + p^2 - 1} + |p|},$$

therefore, the conditions $r_{e,h}(-k,-p) = r_{e,h}(k,p)$ are fulfilled . Also $r_{e,h}(-\omega,p) = r_{e,h}^*(\omega,p)$, $\varepsilon(-k) = \varepsilon^*(k) = \varepsilon(k)$. At the same time $f_\mu(-\xi,-p,d) = f_\mu(\xi,p,d)$. This means that the exponent $\exp(-2mp\xi d/c)$ in DE does not change when the sign of frequency and *p* changes simultaneously, fading out at large d. When switching to frequency $\exp(-2im\omega d/c)$, and taking into account that $2d/c$ is a two-way time for the passage of light through the gap *d* in both directions, we obtain waves $\exp(i\omega(t - 2md/c))$ corresponding to multiple two-way passage of the gap. In this case, the zeros and poles of the functions are symmetrically located in the complex planes $\xi$ and $\omega$ symmetrically relative to both axes. In the $\omega$ plane, symmetry with respect to the imaginary frequency axis means the same imaginary parts and opposite real parts, and symmetry with respect to the real frequency axis means complex conjugate frequencies, i.e., the same real parts and opposite imaginary parts. Therefore, in the $\xi$ plane, symmetry with respect to the real axis $\xi$ means complex conjugate quantities. The integral over $p$ defines integration over the transverse wavenumbers. The case $p$=1 corresponds to a normal drop $r_{e,h}(k,0) = -\left(\sqrt{\varepsilon(k)}-1\right)/\left(\sqrt{\varepsilon(k)}+1\right)$. The $p=\infty$ case corresponds to sliding waves for which $r_e(k,\infty)=-1$, $r_h(k,\infty)=0$.

The transition to an imaginary frequency is widely used in radio engineering as the Laplace transform. In this case, the transfer functions of multipoles and circuits are the ratio of polynomials or meromorphic functions (in the case of distributed circuits) of the variable $\xi = i\omega$. Due to the principles of physical realizability [16], the zeros and poles of dissipative structures lie in the upper half-plane of the complex plane $\omega$ symmetrically relative to the imaginary axis. However, there are zeros and poles in the lower half-plane. They correspond to active environments and structures. When synthesizing filters and similar structures, stable polynomials (Hurwitz polynomials) with zeros in the upper half-plane should be distinguished. The analytical properties of DP and transfer functions are responsible for fulfilling the principles of causality in the propagation of signals in filters, pulses in media, the relationship of induction **D**(**r**,$t$) with the field **E**(**r**,$t$), the relationship of real and imaginary parts of DP by Kramers-Kronig relations [17-19].

In [7,8], the free energy of the structure at zero temperature is defined as

$$E(d)=\frac{\hbar L^2}{8\pi^2}\int_0^\infty \kappa d\kappa \int_{-\infty}^{\infty} d\xi\left[\ln\left(f_e(\xi,\kappa,d)\right)+\ln\left(f_h(\xi,\kappa,d)\right)\right]. \tag{6}$$

Differentiating (6) by $d$, we obtain the specific force

$$F(d)=\frac{\hbar c}{2\pi^2}\int_0^\infty \kappa d\kappa \int_0^\infty d\xi \sqrt{(\xi/c)^2+\kappa^2}\left[\frac{1}{f_e(\xi,\kappa,d)}+\frac{1}{f_h(\xi,\kappa,d)}\right]. \tag{7}$$

Passing to the variable p, we obtain formula (4). When passing to a new variable $x = 2pkd$, formula (4) takes the form

$$F(d)=\frac{\hbar}{32\pi^2 d^4}\int_0^\infty x^3 dx \int_1^\infty \frac{dp}{p^2}\left[\frac{1}{f_e(x,p,d)}+\frac{1}{f_h(x,p,d)}\right].$$

Formulas (6) and (7) are obtained using the complex analysis theorem “the principle of argument” [7,20]. *Theorem* 1. Let the function $f$ be meromorphic in the domain $D$ and its boundary be a continuous curve containing neither zeros nor poles of $f$. Let also $N$ and $P$, respectively, be the number of zeros and poles (taking into account multiplicity) in the domain $D$. Then

$$N - P = \frac{1}{2\pi i}\oint_G \frac{f'(z)}{f(z)}dz. \quad (8)$$

If you do not count the number of zeros and poles, but sum up the zeros and poles themselves, then you should

$$\frac{1}{2\pi i}\oint_G z\frac{f'(z)}{f(z)}dz = \left[\sum_n z_n\right]_{f(z_n)=0} - \left[\sum_n z_n\right]_{f(z_n)=\infty}. \quad (9)$$

The contour is traversed counterclockwise, while the difference between the number of zeros and poles is determined by changing the argument of the function *f* when traversing the contour (*Theorem* 2 of [20]). In our case, the contour follows an imaginary axis and closes along an infinite left semicircle. All zeros and poles lie inside it in the left half-plane and are complex conjugate. If we take a different dependence on time $\xi = -i\omega$ (as in [3.19] with substitution $i \to -i$), then the contour should be closed in the right half-plane. here, zeros are also complex conjugates.

Let us consider the functions $F_{e,h} = (\omega,\kappa) f_{e,h}(\omega,\kappa)\coth(\hbar\omega(\kappa)/(2k_B T))$ in (5). Frequencies $\omega_{n(e,h)}$ are zeros of functions $f_{e,h}(\omega)$. The functions considered also have zeros on the imaginary frequency axis associated with cotangent zeros. They are located at the points $\omega_n^0 = 2ik_B T(n-1/2)\pi/\hbar = i\xi_n^0$, $n = \pm1,\pm2,...$.. For *n*>0, these zeros are in the upper half-plane. In addition to the zeros, the cotangent has poles on the imaginary frequency axis at points $\omega_n^p = 2ik_B Tn\pi/\hbar = i\xi_n^p$, $n = 0,\pm1,\pm2,...$. The imaginary frequencies are the Matsubara frequencies [21]. When rotating $\xi = i\omega$, we represent the cotangent as

$$\cot(z) = \frac{1}{z - z_n} + \psi_n(z),$$

where $\psi_n(z)$ is a regular function, including points $z_n$ in *n* regions $(n-1/2)\pi < z < (n+1/2)\pi$, $n = 0,\pm1,\pm2,...$, $z = k\delta_T/2$, $z_n = n\pi = \xi_n^p \delta_T/(2c)$. We have $\psi_n(-z) = -\psi_{-n}(z)$, $z_{-n} = -z_n$, $\coth(x) = \coth(iz) = -i\cot(iz)$. The contour should not pass through zeros and poles, so they should be bypassed. While Wick rotating $\omega = i\xi = i\xi' - \xi''$, we move on to the half-plane $\xi = \xi' + i\xi''$, Fig. 2. In this case, the plane of the drawing is rotated by an angle of $\pi/2$ for convenience, i.e. the axis of the abscissa is directed along $\xi'$. Note that we get exactly the same pattern if we make a rotation $\omega = -i\xi$, and then rotate the plane of the drawing by an angle $-\pi/2$. In the first case ($\omega = i\xi$), zeros with negative real parts are summed, and in the second case ($\omega = -i\xi$) zeros have positive real parts. The zeros and poles of the cotangent lie on the real axis $\xi' = \mathrm{Im}(\omega)$. They go around the top of Fig. 2 to avoid entering the area. There are zeros and poles of functions $f_{e.h}(\xi, \kappa)$ in the upper half-plane $i\xi'' = -i\omega'$: $\omega_{n(e.h)}^{(0,p)} = \pm\xi_{n(e.h)}'^{(0,p)} - i\xi_{n(e.h)}''^{(0,p)}$. They are symmetrically arranged relative to the axis $\xi'$, i.e., when summing up, their real parts (i.e., the imaginary parts of the frequencies) cancel each other out, while the real parts double. Since "imaginary" quantities in the plane $\xi = \xi' + i\xi''$ correspond to real quantities in the plane $\omega$ and vice versa, namely $\omega'' = \xi'$, $\omega' = -\xi''$, this means that the quantities associated with dissipation are mutually compensated, and the quantities $-i\xi_{n(e.h)}''^{(0,p)} = i\omega_{n(e.h)}'^{(0,p)}$ are summed. Having a multiplier $1/(2\pi i)$ in front of the contour integral makes the sum valid and positive. This explains why the final expressions are valid when dealing with complex quantities. Bypassing the poles leads to the appearance of semi–residuals in the points $\xi_n^p(\kappa)$, and the integral with the punctured points should, generally speaking, be understood in the sense of p.v. - the principal value according to Cauchy. For zeros, integral in the sense of the p.v. is the usual integral, so their bypass is not shown in the figure. We will take the sum of the half-residuals for positive values, doubling it, and take the half- residuals at zero with weight of ½. The sum (3) determines the free energy and is multiplied by the sum of the zeros of the considered functions. They depend on *d*. The poles of these functions do not depend on *d*, since they are determined from the conditions $r_{e,h}^2 = 0$ and give a constant contribution of *C*. When differentiating by *d*, it disappears. Therefore, we will omit it. We have the result:

$$E(d) = \frac{\hbar L^2}{4\pi(2\pi i)} \int_0^\infty \kappa d\kappa \sum_{\mu=e,h} \oint_G \omega \frac{F'_\mu(\omega,\kappa,d)}{F_\mu(\omega,\kappa,d)} d\omega + E_p(d). \tag{10}$$

Here $E_p(d)$ is the contribution to the energy from the deductions at the poles of the cotangent, which is given by the function $1/(z - z_n)$. A contour integral can be associated with the function $\psi_n(z)$. It should be noted that *d*-independent and, generally speaking, infinite contributions can be associated with the infinite energy of the quantum vacuum outside the structure of Fig, $D \to \infty$. Turning to the counter in the plane $\xi$, we see that when summing the frequencies, the imaginary parts of them are compensated, and the real parts are doubled. For energy, a real result is obtained. The integral along an infinitely large right or left semicircle in the plane $\xi$ disappears, and the integral along the imaginary axis remains. For it, we have (see [7])

$$\int_{-\infty}^{\infty}(-i\xi)\frac{\partial F_\mu(-i\xi)/\partial(-i\xi)}{F_\mu(-i\xi)}(-id\xi)=$$

$$=-i\int_{-\infty}^{\infty}\frac{\partial_\xi F_\mu(-i\xi)}{F_\mu(-i\xi)}\xi d\xi= \quad (11)$$

$$=-i\int_{-\infty}^{\infty}\xi d\xi\frac{d}{d\xi}\ln(F_\mu(-i\xi))=$$

$$=i\int_{-\infty}^{\infty}\ln(F_\mu(-i\xi))d\xi\,.$$

Here we have integrated in parts, and omitted the infinite substitutions that do not depend on *d* and disappear during differentiation. In [7], to justify the disappearance of the integral over an infinite semicircle, it is stated that it does not depend on *d*, therefore it should be omitted. However, its disappearance follows from the fact that there is a large positive value of $\xi$ on the infinite semicircle, zeroing out the exponent. The values $f_\mu(-i\xi)=f_\mu(\omega)$ are denoted by the same symbols, i.e. $f_\mu(i\xi,\kappa,d)=f_\mu(\xi,\kappa,d)$. After differentiating the logarithm, we obtain a term independent of *d* and the structure parameters, which we omit [7]. The *d*-independent terms can be explained by virtual and thermal photons from the outside of the structure in Fig. 1, whose energy and pressure are infinite, and their difference with the energy and pressure of photons inside the gap gives finite and *d*-dependent results. The force in the form of external Casimir pressure $P=F=F_L(d,T)+\tilde{F}(d,\delta_T)$ is obtained by differentiating in *d*, while the terms $(z-z_n)^{-1}$ in the cotangent give residuals and determine the term $F_L(d,T)$. It is given by Lifshitz's formula, given by him in [10] in the form

$$F_L(d,\delta_T)=\frac{k_B T}{\pi c^3}\sum_{n=0}^{\infty}{}' \xi_n^3\int_1^{\infty}\sum_{\mu=e,h}f_\mu^{-1}(\xi_n,p)p^2dp\,. \quad (12)$$

The second term is related to the contour integral. When replaced $z=k\delta_T/2$ and by extending integration to negative frequencies, it is equal to

$$\tilde{F}(d,\delta_T)=-\frac{4\hbar c}{\pi^2\delta_T^4}\,p.v.\int_{-\infty}^{\infty}z^2|z|\cot(z)\int_1^{\infty}\sum_{\mu=e,h}\frac{r_\mu^2(z,p)\exp(-4p|z|d/\delta_T)}{1-r_\mu^2(z,p)\exp(-4p|z|d/\delta_T)}p^2dpdz\,. \quad (13)$$

Since $\cot(z)=\psi_n(z)+1/(z-z_n)$, the last term has already been taken into account in (12), and the substitution gives the integral (13) in the usual sense.

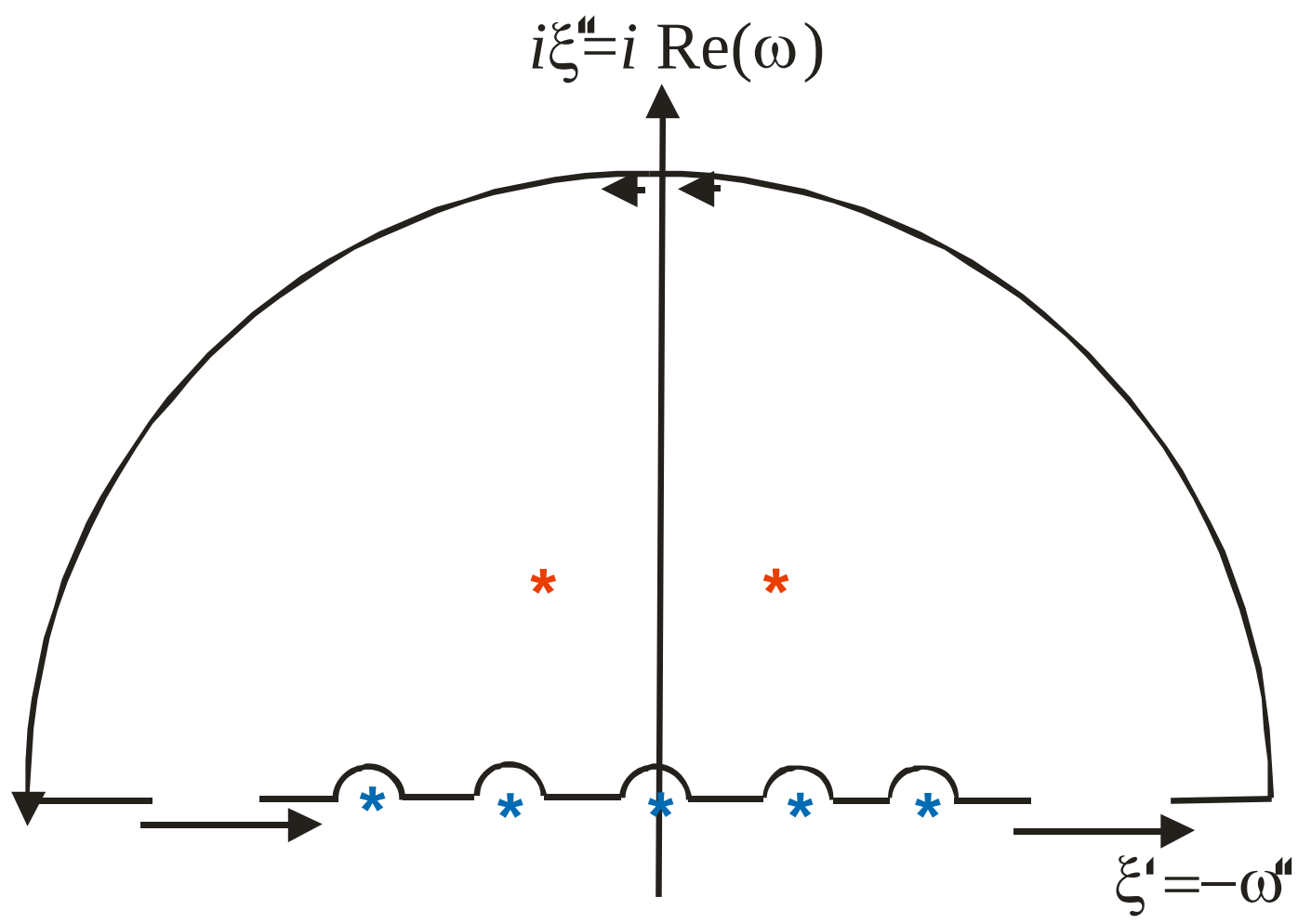

Fig. 2. Outline for the argument principle theorem. The symbols “*” schematically represent the poles of the hyperbolic cotangent, and the symbols “*” represent the zeros of the functions $f_{e,h}$

Lifshitz in [10] twice justified the absence of a contour integral by taking the real part in the initial formula given by him and by the fact that the values of the hyperbolic cotangent on the imaginary axis are imaginary. However, in this formula, the cotangent is included with the multiplier $\omega^3$, so the result is real. At the same time $d\omega = -id\xi$, with a multiplier $1/(2\pi i)$, which also give a real result.

Consider formula (12) for $r_{e,h} = -1$. Using the Neumann series expansion, performing integration with respect to *p*, denoting $x = 4\pi d / \delta_T$ and highlighting the term *n*=0, we obtain

$$F_L(d,T) = \frac{\hbar c}{4\pi d^3 \delta_T}\left[\varsigma(3) + \sum_{m=1}^{\infty}\sum_{n=1}^{\infty} \exp(-xnm)\left(\frac{x^2 n^2}{m} + \frac{2xn}{m^2} + \frac{2}{m^3}\right)\right]. \quad (14)$$

At a high temperature (*x*<<1), we obtain

$$F_L(d,T) \approx \frac{\hbar c}{4\pi d^3 \delta_T}\left[\varsigma(3) + \exp(-x)\left(x^2 + 2x + 2\right)\right] \approx \frac{\hbar c \varsigma(3)}{4\pi d^3 \delta_T},$$

This is consistent with the results of [2,3,22], however, the result [22] was obtained on the basis of free energy determined by the Lifshitz formula. At low temperatures (*x*>>1), the m index runs extremely close to the values. Replacing the sum with an integral with a substitution $dm = dz/(xn)$, we get with integration by parts

$$\sum_{m=1}^{\infty} dm = (nx)^2 \int_{nx}^{\infty} \exp(-z)\left(\frac{1}{z}+\frac{2}{z^2}+\frac{2}{z^3}\right)dz = \\ = \frac{\exp(-nx)}{nx}+\frac{\exp(-nx)}{(nx)^2} .$$

The result is also easily verified by *nx* differentiation. Performing summation over *n* by integration over $dn = dz/x$, we find after integration in parts

$$\int_{x}^{\infty} \exp(-z)\left(\frac{1}{z}+\frac{1}{z^2}\right)dz = \frac{\exp(-x)}{x} .$$

The result is checked by differentiating by *x*. We get the result according to the Lifshitz formula, which is about 6.5 times less than that of Casimir:

$$F_L(d,T) = \frac{\hbar c}{4\pi d^3 \delta_T}\left[\varsigma(3)+\frac{\exp(-4\pi d/\delta_T)}{4\pi d/\delta_T}\right] \approx \frac{\hbar c}{16\pi^2 d^4} \approx 0.154 F_C . \tag{15}$$

Consider the following asymptotic summation formulas for $x \to 0$. We have for $\alpha \geq 0$, $y = mx$, $dm = dy/x$:

$$\sum_{m=1}^{\infty} m^{\alpha} \exp(-mx) = \frac{1}{x^{\alpha+1}} \int_{x}^{\infty} y^{\alpha} \exp(-y) dy =$$

$$= \exp(-x)\left[\frac{1}{x}+\frac{\alpha}{x^2}+\ldots+\frac{\alpha!}{x^{\alpha}}+\frac{\alpha!}{x^{\alpha+1}}\right] \sim$$

$$\sim \frac{\alpha!}{x^{\alpha+1}} \exp(-x) \sim \frac{\alpha!}{x^{\alpha+1}} .$$

These formulas are obtained by integrating the exponential parts with disregard for the remaining small terms and give the correct law of increasing sums for divergent series. We also have the formula

$$\sum_{m=1}^{\infty} m^{-1} \exp(-mx) = \int_{x}^{\infty} y^{-1} \exp(-y) dy = \\ = -\mathrm{Ei}(-x) \sim \ln(x) .$$

The specified series at $x$=0 diverges logarithmically. There are also formulas for convergent series

$$\sum_{m=1}^{\infty} \frac{\exp(-mx)}{m^{\alpha}} = \varsigma(\alpha).$$

Using the above formulas, we obtain after summation in (14) by $n$

$$F_L(d,T) = \frac{\hbar c}{4\pi d^3 \delta_T} \left[ \varsigma(3) + \frac{6}{x} \sum_{m=1}^{\infty} \exp(-xm) \frac{1}{m^4} \right].$$

The series in $m$ converges, and we get

$$F_L(d,T) = \frac{\hbar c}{4\pi d^3 \delta_T} \left( \varsigma(3) + 6\varsigma(4)/x \right) \approx \frac{3\hbar c \varsigma(4)}{8\pi^2 d^4}.$$

This result is also less than FC, but unlike (15), it is obtained in excess, since the series diverge at $x$=0 in the reverse order of summation. We have found that the interaction force of ideal screens at short distances according to the Lifshitz formula (12), although it has an asymptotic value $1/d^4$, is less than the result of Casimir and the result [2]. In order to match Casimir's result, the double sum in (14) would have to have asymptotic $\pi^3 \delta_T /(60d)$. Although (12) at $T \to 0$ and for finite $d$ formally implies (4) by replacing the sum with an integral, but at $r_{\mu}^{e} = 1$, finite temperature and $d \to 0$ it does not follow the result of (2) and the Casimir's result. This suggests that for small d, the Lifshitz formula (12) is not exact.

Due to the periodicity of the cotangent

$$\cot(z) = \sum_{n=-\infty}^{\infty} \frac{1}{z - n\pi}$$

formula (13) can be reduced to an integral in a finite domain, and for the ideal case $r_{e,h} = -1$ consider the analytical calculation of the integral. For dissipative structures with dispersion and $r_{e,h}^2(z,p) < 1$ it is also quite laborious, so it is more convenient to calculate (13) numerically. Let's define the area $n$ as $(n-1/2)\pi < z < (n-1/2)\pi$. In this area $\psi(z) = \psi_n(z)$,

$$\psi_n(z) = \sum_{\substack{k=-\infty \\ k \neq n}}^{\infty} \frac{1}{z - k\pi} = \cot(z) - \frac{1}{z - n\pi}.$$

We will also define $\psi_{-n}(z) = -\psi_n(-z) = -\psi(z)$ in the field the area $-n$. Then $\psi_n(z) - \psi_{-n}(z) = 1/(z + n\pi) + 1/(z - n\pi)$. Considering even functions of *z*, the integral can be reduced to the domain $(0, \infty)$:

$$\tilde{F}(d, \delta_T) = -\frac{8\hbar c}{\pi^2 \delta_T^4} \int_0^\infty z^3 \psi(z) \sum_{\substack{m=1 \\ \mu=e,h}}^\infty \int_1^\infty r_\mu^{2m}(z, p) p^2 \exp(-a_m z p) dp dz, \tag{16}$$

where $a_m = 4md/\delta_T$. The region *n*=0 is taken into account in integral (13) once, whereas the regions *n* > 0 and n < 0 are taken into account twice. Expression (16) is obtained by expanding the fractions in (12) into the Neumann series.

For $T \to 0$ and finite *d* we have $\delta_T \to 0$, and the contribution (16) disappears. Indeed, with $r_\mu^2 = 1$ the field of integration with respect to *p*, we obtain a large term proportional to $1/a_m^3 \sim \delta_T^3$. Considering the multiplier for integral (16), we get zero in the limit. This can also be seen after integration over *p*. The maximum contribution is given by a member $\exp(-a_m z)/a_m$. After integrating over *z*, taking into account the periodicity, we obtain a decaying multiplier $(\delta_T / a_m d)^2 \exp(-2\pi n m d/\delta_T)/n$. If the temperature is not zero, but $d \to 0$, then after integration over *p*, the maximum contribution is given by the term with $1/a_m \sim \delta_T/d$. In this case (16) gives a contribution tending to infinity. As for the Lifshitz formula (12), then at $T \to 0$ and for finite *d* it passes into formula (1). However, if the temperature is not zero, but $d \to 0$, it does not correspond to (2) and (4), and the result (14) occurs, which does not correspond to (1).

It is convenient to reduce the integral (13) to the domain $(0, \pi/2)$. We denote the *z* -dependent integrand function as $\psi(z) r_\mu^{2m}(z, p) \varphi_{\mu m}(z, p)$, where $\varphi_m(z, p) = z^3 \exp(-a_m p|z|)$. When $r_\mu^2 = 1$ the integral of the function $\varphi_m(z, p)$ with respect to *p* is denoted by $\Phi_m(z)$. It converges by virtue of a decreasing exponent. The integral of the function $z^3 \exp(-a_m |z| p)$ with respect to *p* is

$$\Phi_m(z) = \exp(-a_m|z|)\left[\frac{z^2\,\mathrm{sgn}(z)}{a_m} + \frac{2z}{a_m^2} + \frac{2\,\mathrm{sgn}(z)}{a_m^3}\right].$$

In the case of non-ideal structures, it should be decomposed $r_\mu^{2m}(z,p)$ in z and *p*, then performing integration in *p*. By swapping variables $z' = z \mp n\pi$, we reduce both integrals over the specified regions $(-\pi/2, \pi/2)$ to an integral over the region $(-\pi/2, \pi/2)$ with weight 2 for positive *n* and with weight 1 for *n*=0. Summing up, we get an integral over *z* as an infinite sum of *n* integrals over the domain $(-\pi/2, \pi/2)$. We reduce them to an area $(0, \pi/2)$. Then the first term of the sum for *n*=0 should be taken with a weight of 1/2, and the rest with a unit weight in the same way as in the Lifshitz formula. The resulting series converges because the index n is included in the decaying exponent. The form of formula (16) shows why Lifshitz's formula is very accurate at a finite (high) temperature. Recall that the room temperature is considered high at *d* of the order of and more than 1 microns. In this case, the exponent becomes very small, and the contribution (16) to the force is small compared to the contribution from the Lifshitz formula. This case implements almost all applications, with the exception of very small *d*. As *d* tends to infinity at any finite temperature, (16) also tends to zero. At a finite temperature and $d \to 0$, both integrals (12) and (16) diverge, as do (3). In an Ideal case $r_{e,h}^2 = 1$, they should sum up to the Casimir result with $d \to 0$, which in this case tends to infinity. We see that in this case, when striving $d/\delta_T \to 0$, the formula (12) behaves differently with respect to (4), depending on whether the temperature or distance is finite. At a finite distance and an infinitesimal temperature $d/\delta_T \to 0$, i.e. $\delta_T \to \infty$, and at (12), formula (4) follows, giving the Casimir result. However, if the temperature is finite, aad $d/\delta_T \to 0$, then the formula (12) does not change to (4).

It is easiest to investigate formulas (13) and (16) in the ideal case $r_\mu^2 = 1$. Then numerical estimates of formula (16) show that the real difference from the Lifshitz formula at room temperature occurs at $d$ less than 100 nm. At the same time, the contribution of (16) is positive, i.e. the Lifshitz formula, as shown in Fig. 3, gives underestimated values. The contribution (16) was calculated at $d$=10 nm, $T$=300 K and was $0.094 \cdot 10^6$ N/m$^2$. In this case, formula (12) gives the value $0.0199 \cdot 10^6$ N/m$^2$. Their sum is close to Casimir's result of $0.1295385 \cdot 10^6$ N/m2. Consider the analytical estimates in this case. Let's denote the integral over the domain $n$

$$I_n(a_m) = \int_{-\pi/2+n\pi}^{\pi/2+n\pi} z^3 \left[\cot(z) - \frac{1}{z - n\pi}\right] \int_1^{\infty} p^2 \exp(-a_m |z| p) dp dz. \tag{17}$$

After replacing ($n > 0$) and omitting the stroke, we get

$$I_n(a_m) = \int_{-\pi/2}^{\pi/2} (z + n\pi)^3 \left(\cot(z) - \frac{1}{z}\right) \int_1^{\infty} p^2 \exp(-a_m (z + n\pi) p) dp dz. \tag{18}$$

We have an approximation $\cot(z) - 1/z \approx \psi_0 = -z/3 - z^3/45 - 2z^5/945$ [23]. You can take it as $\psi_0 = -4z/\pi^2$. With this approximation $\cot(\pi/2) = 0$. However, instead of 1, it gives $\cot(\pi/4) \approx 0.955$ with an error of about 5%. Reducing the integral to the domain $(0, \pi/2)$, we have

$$I_n(a_m) = \int_1^{\infty} \tilde{I}_n(a_m p) p^2 \exp(-a_m n\pi p) dp,$$

$$\tilde{I}_n(a_m p) = \int_0^{\pi/2} \varphi_n(z, a_m, p) dz, \tag{19}$$

$$\varphi_n(z, a_m, p) = \psi_0(z) \left[(z + n\pi)^3 \exp(-a_m z p) + (z - n\pi)^3 \exp(a_m z p)\right].$$

We 'll take it as $\psi_0 = -4z/\pi^2$. Then $\tilde{I}_n(a_m p) = \tilde{\tilde{I}}_n(a_m p) + \tilde{\tilde{I}}_{-n}(-a_m p)$ where

$$\tilde{\tilde{I}}_n(a_m p) = \frac{4}{\pi^2} \int_0^{\pi/2} \left(z^4 + 3z^3 n\pi + 3z^2 (n\pi)^2 + z(n\pi)^3\right) \exp(-a_m p z) dz.$$

This integral is calculated using the well-known formulas [23], which are a rather cumbersome result. Therefore, we will use the approximation of small and large $a_m$. Let $a_m >> 1$, which corresponds to the case of high temperature $d/\delta_T >> 1$. Since $a_m$ it appears in the denominator to a power one greater than the power of *z*, we take the last term with the smallest degree, and we get the contribution

$$\tilde{\tilde{I}}_{n1}(a_m, p) \approx \frac{4}{\pi^2}(n\pi)^3 \left\{ \frac{1}{(a_m p)^2} - \exp\left(-\frac{a_m \pi p}{2}\right)\left[\frac{\pi/2}{a_m p} + \frac{1}{(a_m p)^2}\right]\right\}.$$

The contribution of this integral to $\tilde{I}_n(a_m p)$ is

$$\tilde{I}_{n1}(a_m p) = -\frac{4}{\pi^2}(n\pi)^3 \exp\left(-\frac{a_m \pi p}{2}\right)\left[\frac{\pi/2}{a_m p} + \frac{1}{(a_m p)^2}\right] -$$

$$+\frac{4}{\pi^2}(n\pi)^3 \exp\left(\frac{a_m \pi p}{2}\right)\left[-\frac{\pi/2}{a_m p} + \frac{1}{(a_m p)^2}\right] \approx$$

$$\approx -\frac{4}{\pi^2}(n\pi)^3 \exp\left(\frac{a_m \pi p}{2}\right)\frac{\pi/2}{a_m p}.$$

Given a term with degree 2, we get

$$\tilde{I}_{n2}(a_m p) = \tilde{\tilde{I}}_{n2}(a_m p) + \tilde{\tilde{I}}_{-n2}(-a_m p) \approx$$

$$\approx \frac{12(n\pi)^2}{\pi^2}\int_0^{\pi/2} z^2 \exp(a_m pz)dz \approx \frac{12(n\pi)^2}{\pi^2}\frac{(\pi/2)^2 \exp(a_m pz)}{(a_m p)}.$$

Similarly for degree 3

$$\tilde{I}_{n3}(a_m p) = \tilde{\tilde{I}}_{n3}(a_m p) + \tilde{\tilde{I}}_{-n3}(-a_m p) \approx$$

$$\approx -\frac{12n\pi}{\pi^2}\int_0^{\pi/2} z^3 \exp(a_m pz)dz \approx -\frac{12n\pi}{\pi^2}\frac{(\pi/2)^3 \exp(a_m pz)}{(a_m p)},$$

$$\tilde{I}_{n4}(a_m p) \approx \frac{4}{\pi^2}\int_0^{\pi/2} z^4 \exp(a_m pz)dz \approx \frac{4(\pi/2)^4}{\pi^2}\frac{\exp(a_m p\pi/2)}{a_m p}.$$

$$\tilde{I}_n(a_m p) \approx 4\pi^2 \frac{\exp(a_m p\pi/2)}{a_m p}\left[\frac{1}{2^4} + \frac{3n}{2^3} + \frac{3n^2}{2^2} + \frac{n^3}{2}\right] \approx$$

$$\approx 2\pi^2 n^3 \frac{\exp(a_m p\pi/2)}{a_m p},$$

$$I_n(a_m) = \frac{2\pi^2 n^3}{a_m}\int_1^\infty p\exp(-a_m(n-1/2)\pi p)dp.$$

After integration by $p$, we get

$$I_n(a_m) \approx 2\pi^2 n^3 \frac{\exp(-a_m(n-1/2)\pi)}{a_m^2(n-1/2)\pi}.$$

The full contribution of the contour integral is

$$\tilde{F}(d,\delta_T) = \frac{16\hbar c}{\pi^2\delta_T^4}\sum_{m=1}^\infty\sum_{n=1}^\infty I_n(a_m) =$$

$$\approx \frac{2\hbar c}{\pi^3\delta_T^2 d^2}\sum_{m=1}^\infty\sum_{n=1}^\infty \frac{n^3}{m^2(n-1/2)}\exp\left(-\frac{4m(n-1/2)\pi^2 d}{\delta_T}\right) \approx$$

$$\approx \frac{4\hbar c}{\pi^3\delta_T^2 d^2}\exp\left(-\frac{2\pi^2 d}{\delta_T}\right).$$

Here we doubled the result due to negative n and doubled it again due to the same contribution of E-modes and H-modes. We took one term each in the sums, since the exponents are extremely small. This positive contribution to the force is very small compared to the contribution from the Lifshitz formula. It is possible to determine the boundary of the distance when the contribution (13) is not significant compared to the contribution from the Lifshitz formula: $\tilde{F}/F_L = 16(d/\delta_T)\exp(-2\pi^2 d/\delta_T)d/\delta_T$. We have $\tilde{F}/F_L = 0.108$ already at $d/\delta_T = 0.01$. For room temperature, this is 76 nm. For $d > 100$ nm, the formula (2) can be considered quite accurate. The Lifshitz formula is very accurate for $d/\delta_T >> 1$. At a fixed temperature (i.e. $\delta_T$), this case is realized at $d \to \infty$.

Let's consider another limiting case $d/\delta_T << 1$. At a fixed temperature it is realized at $d \to 0$, and at a fixed distance at $T \to 0$. For ideally adducting screens, this ideal case does not affect the need to limit distances by atomic dimensions and may well be considered. It is implemented at $a_m = 4md/\delta_T << 1$. Therefore, a limitation should be made in the sum of *m*, i.e., as *d* decreases, the upper limit of summation should be increased proportionally to $\delta_T/(4d)$. In this case, consider the contribution to (16)

$$\tilde{F}_n(d,\delta_T) = -\frac{8\hbar c}{\pi^2\delta_T^4}\, p.v.\int_{(n-1/2)\pi}^{(n+1/2)\pi} z^3 \cot(z) \sum_{m=1}^{\infty}\int_1^{\infty} p^2 \exp(-a_m zp) dp dz$$

for area *n*: $((n-1/2)\pi, (n+1/2)\pi)$, $n = 0, \pm1, \pm2, \ldots$. In the region of *n*=0, we can take an approximation $\cot(z) = 1/z - z/3 - z^3/45 - \ldots$. However, we will take it for convenience $\cot(z) = 1/z - 4z/\pi^2$. Explicitly selecting the cotangent pole, we obtain

$$\tilde{F}_n(d,\delta_T) = -\frac{8\hbar c}{\pi^2\delta_T^4}\int_{(n-1/2)\pi}^{(n+1/2)\pi} z^3\left[\cot(z) - \frac{1}{z - n\pi}\right]\sum_{m=1}^{\infty}\int_1^{\infty} p^2 \exp(-a_m zp) dp dz .$$

This integral can be understood in the usual sense. The full contribution to the contour integral is

$$\tilde{F}(d,\delta_T) = \sum_{n=0}^{\infty}{}' \tilde{F}_n(d,\delta_T). \tag{20}$$

We take the zero term with a weight of 1/2. In the area of *n* we replace $\cot(z) - (z - n\pi)^{-1} \approx -4(z - n\pi)/\pi^2$. After integration by *p*, we have

$$\tilde{F}_n(d,\delta_T) = \frac{\hbar c}{2\pi^2\delta_T d^3}\int_{(n-1/2)\pi}^{(n+1/2)\pi} (z - n\pi) S(z, d/\delta) dz , \tag{21}$$

$$S(z, d/\delta) = \sum_{m=1}^{\infty}\exp\left(-\frac{4mzd}{\delta_T}\right)\left[\frac{z^2(4d/\delta)^2}{m} + \frac{2z(4d/\delta)}{m^2} + \frac{2}{m^3}\right].$$

After replacing $z = z' - n\pi$, summing over the regions of *n* and omitting the stroke, we obtain an integral over the region $(-\pi/2, \pi/2)$, which, by virtue of parity, leads to the region $(0, \pi/2)$:

$$I_{mn}(d/\delta_T) = 2\int_0^{\pi/2} z \exp(-mzx)\left[\frac{(z+n)\pi^2 x^2}{m} + \frac{2(z+n)x}{m^2} + \frac{2}{m^3}\right] dz . \tag{23}$$

It indicates an infinitesimal value $x = 4d/\delta_T$. The contribution to the force takes the form

$$\widetilde{F}(d,\delta_T) = \frac{\hbar c}{2\pi^2 \delta_T d^3} \sum_{\substack{m=1\\n=0}}^{\infty}{}' \exp\left(-\frac{4n\pi d}{\delta_T}\right) I_{mn}(\delta_T/\delta_T). \tag{24}$$

When calculating (23), we consider the integrals

$$I_{m1}(x) = \int_0^{\pi/2} z\exp(-mzx)dz = -\exp(-mxz)\left[\frac{z}{mx} + \frac{1}{(mx)^2}\right]_{z=0}^{z=\pi/2}.$$

$$I_{m2}(x) = \int_0^{\pi/2} z^2\exp(-mzx)dz = -\exp(-mxz)\left[\frac{z^2}{mx} + \frac{2z}{(mx)^2} + \frac{2}{(mx)^3}\right]_{z=0}^{z=\pi/2}.$$

Decomposing by a small parameter *x*, we obtain

$$I_{m1}(x) \approx \frac{\pi^2}{8} - \frac{\pi^3 mx}{16},$$

$$I_{m2}(x) \approx \frac{\pi^3}{24} - \frac{\pi^4 mx}{64} + \frac{\pi^5 (mx)^2}{320}.$$

As a result, we get

$$I_{mn}(d/\delta_T) \approx \frac{\pi^2}{2}\frac{1}{m^3} + \frac{x}{m^2}\left[\frac{\pi^3}{6} + \frac{\pi^2}{2}n - \frac{\pi^3}{4}\right] +$$

$$+\left[\pi^4\left(\frac{\pi}{12} - \frac{1}{16}\right) + \frac{\pi^3(\pi-1)}{4}n\right]\frac{x^2}{m} -$$

$$-\pi^5 x^3\left(\frac{\pi}{32} - \frac{1}{80} + \frac{n}{8}\right) + \frac{\pi^7 mx^4}{160}.$$

When summing, you should omit the small terms, i.e. take:

$$I_{mn}(d/\delta_T) \approx \frac{\pi^2}{2}\left(\frac{1}{m^3} + \frac{xn}{m^2}\right). \tag{25}$$

We see that there is an additional positive contribution to the Lifshitz result (14). To obtain accurate results, formulas (14) and (24) should be numerically summarized. The series can be calculated with any accuracy, given that the number of terms in the sums should be increased proportionally to $1/x$ and the convergence should be numerically evaluated for fixed $d$. Such calculations for (14) are shown in Fig. 3. It can be seen from the figure that formula (14) correctly describes the force at $d >$ 1 μm. At $d <$ 100 nm, the results of (14) are much lower than the Casimir result, which should correspond to the force at such small distances according to [2]. It should be noted that formula (2) gives the contribution of the second term more than the contribution of the first one already at $d/\delta_T = 1$, i.e. the temperature correction to the Casimir force is positive. At room temperature $d/\delta_T = 0.4386 \cdot 10^{-4}$ and $d$ =10 nm, we have $2 \cdot 10^{-17}$, i.e. the correction is practically zero, i.e. taking into account the first term with the exponent in (2) practically makes no sense. It is advisable to estimate the contributions of (14) and (24) using the asymptotic summation formulas. The result (15) is obtained by summing first over m and then over n. Summing in reverse order, we obtain

$$F_L(d,T) = \frac{\hbar c}{4\pi d^3 \delta_T}\left[\frac{4\varsigma(4)}{x} + 7\varsigma(3) + x\varsigma(2)\right] \approx \frac{\hbar c \varsigma(4)}{4\pi^2 d^4} = 0.666 P_C .$$

This result should be considered as an upper bound, since the expressions that diverge at $x \to 0$ are asymptotically summed, unlike the summation in the first case. We use (25) for the contribution of (24). Summing first over $n$, then over $m$, we obtain up to small terms

$$\tilde{F} = \frac{\hbar c \pi^2 \varsigma(4)}{8\pi^3 d^4 4} = \frac{\hbar c \pi^2}{229.188 d^4} = 1.047 F_C . \quad (26)$$

This result was obtained in excess. In addition, we approximated the function $\psi_n$ with an excess. Taking it as $\psi = -x/3$, we get

$$\tilde{F} = \frac{\hbar c \pi^2 \varsigma(4)}{8\pi^3 d^4 4} = \frac{\hbar c \pi^2}{229.188 d^4} = 0.8613 F_C . \quad (27)$$

Together with (15) we obtain $F_L + \tilde{F} = 1.015 F_C$. Thus, it is shown that, in sum, the Lifshitz formula and the contour integral give the Casimir result. It is easy to obtain the decomposition of the force over a small parameter *x*, which we do not give now.

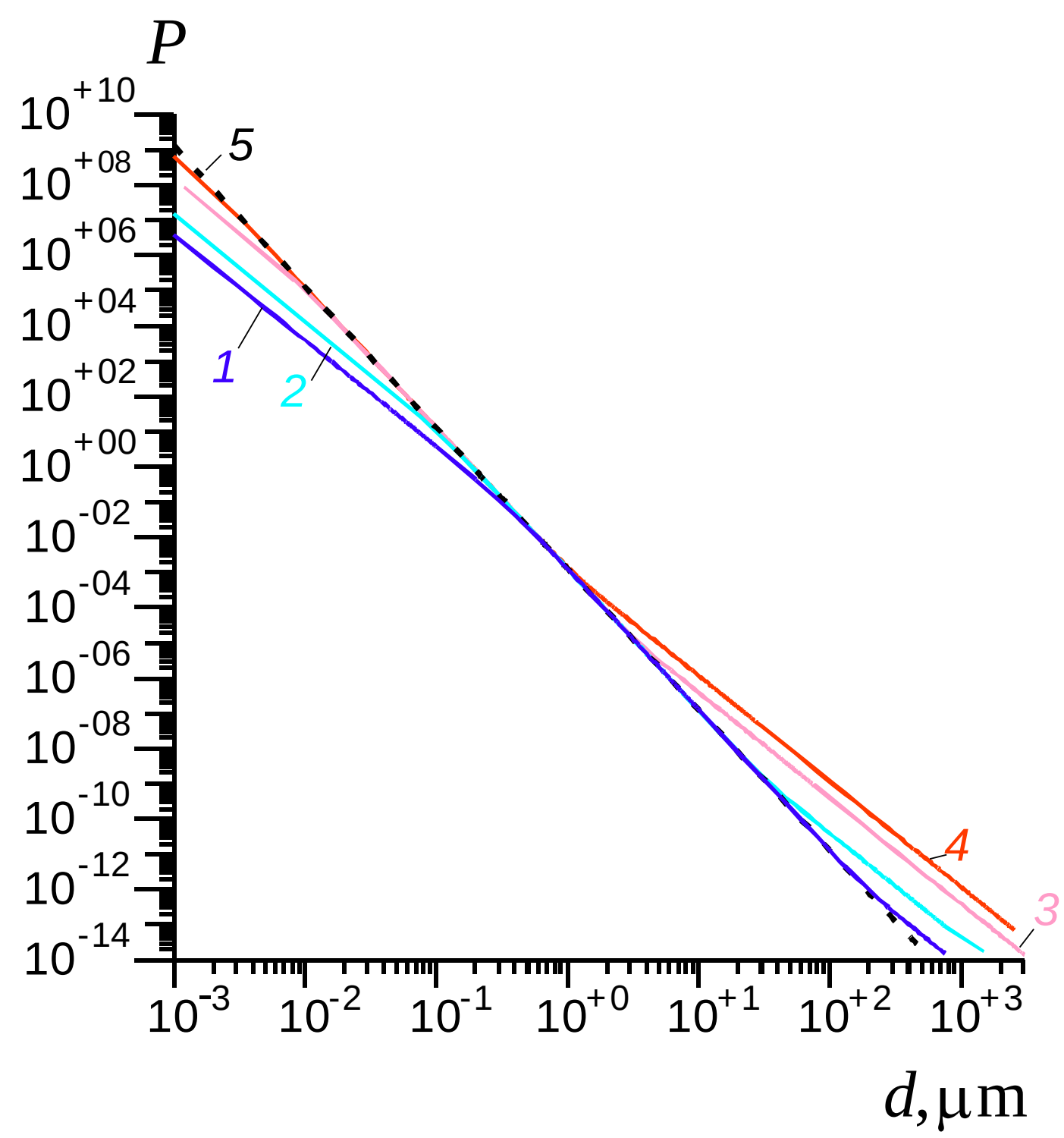

3. External Casimir pressure $P = F$ (H/m$^2$) on two ideally conductive metal layers depending on *d* (in μm) at different temperatures according to formula (14): 5 K (curve 1), 30 K (2), 300 K (3) and 900 K (4). Casimir's result is a dashed line 5

There is no dissipation in the Casimir problem, all frequencies are valid. In real structures, the frequencies are complex, so the possibility of obtaining free energy by summing complex frequencies has given rise to a discussion [7,8-10], since energy cannot be complex. In [10], to get around this, Lifshitz took the real part. The intermediate relation (2.2) given by him in [10] diverges when integrated along the real frequency axis, as does the final relation (2.4), in which the real part is also taken. Nevertheless, it is considered an integral of the analytic function *p*. The expression (8.60) for the Lifshitz formula from [7] also diverges when integrated along the real frequency axis. To circumvent this difficulty, then the rotation of the Wick $\omega = -i\xi$ is used. Such a Lifshitz formula after rotation is a real expression without divergences and does not contain taking the real part. Note that formula (8.60) in [7] is complex, does not contain a real part, and is justified, as in [3], by the development of source theory. Divergence is usually associated with the infinite energy of the quantum vacuum. The question arises: is it necessary to take the real part, and how does the summation of complex frequencies lead to real values for the force? It is shown above that in the framework of the argument-based principle theorem approach, this is easy to understand, since the imaginary frequencies are compensated during summation. From the point of view of electrodynamics, it is also easy to understand. The effect of a monochromatic wave on a dissipative medium means that the losses over the period are exactly compensated by the field source, i.e. the amplitudes do not decrease, although the medium heats up. In weak fields and with high heat capacity, the temperature can be considered constant. Heating leads to the distribution of energy over the entire spectrum of the equilibrium Planck field, i.e. corresponds to the loss of energy of the monochromatic wave, which is fully compensated by the source. The problem of determining the free energy in this case boils down to the fact that the stored energy in the medium is not known [19, 24]. It can, for example, consist in the kinetic energy of particle motion, for example, for plasma. If it is taken into account, the free energy can be determined explicitly [24]. For dispersion forces, fluctuations are as the sources of fields. The energy absorbed by the medium is distributed over the entire equilibrium Planck spectrum. However, according to the Kirchhoff principle, a body at each frequency emits exactly as much as it absorbs, i.e. there are no losses. Another question also arises: the Lifshitz formula (12) was proved in [25,26] and in [3], and its correctness is beyond doubt. Consider these proofs. In [25,26], Lifshitz and his co-authors constructed a macroscopic theory of Casimir - van der Waals forces based on the methods of quantum field theory in statistical physics and obtained the proof of formula (12) using Matsubara's approach [18] to quantum statistical physics, which consists in introducing imaginary time instead of temperature, varying in finite limits from zero to reverse temperature: $0<\tau<1/(k_B T)$. The equations of quantum field theory do not formally change and allow the use of perturbation theory and diagrammatic techniques. In fact, it is possible to construct polarization operators (Green functions) using Feynman diagrams. The thermodynamic properties of the system are determined by the statistical sum or matrix $\rho=\exp(-H/(k_B T))$ associated with the Hamiltonian *H*, through which the free energy is expressed using the relation $E=-k_B T\ln(Sp(\rho))$. The theory is based on the following approximation: "*the distances between the surfaces of bodies are assumed to be small, but still large compared to the interatomic distances in bodies*" [26], which implies the use of perturbation theory, i.e. the representation of the Hamiltonian $H = H_0 + \delta H$ in the form of the main part $H_0$ and the perturbation $\delta H$ associated with "*long-wave electromagnetic fluctuations*" [26]. In theory, summation over the Matsubara frequencies immediately arises. At the same time, it is impossible to go directly to the zero temperature limit: the value $1/(k_B T)$ is included in all the basic equations and uncertainty arises. The limit $T \to 0$ means that there are no thermal fluctuations, i.e., the solution of the problem of the interaction of atoms according to van der Waals with the Hamiltonian $H_0$. Taking into account only long-wavelength electromagnetic fluctuations means that at distances of at least the order of 1–10 nm, the theory is not applicable. A note was made in [26]: "*The allocation of the long-wavelength part mathematically means that the integral in this formula is cut off in some way at small distances. However, we will not explicitly introduce this circumcision, since the answer does not depend on it.*" However, this response does not depend on large distances, but on small ones of the order of 1–10 nm it depends and significantly. It should be noted that the Matsubara frequencies were introduced by Lifshitz actually before the publication of Matsubara's own work. As for Schwinger's proof, he derived formula (4) for $T$=0 from the variational principle, and simply made a substitution to account for temperature (see formula (2.34) from [3]) $\varsigma \to \varsigma_n = 2\pi n/(k_B T)$:

$$\int_0^\infty (\;)\frac{d\varsigma}{2\pi} \to \frac{1}{k_B T}\sum_{n=0}^{\infty}(\;)' \;.$$

(the variable $\zeta$ is our value $\xi$). At the same time, for high temperatures and for ideally conducting plates, the main contribution is given by the term with $n$=0, and the result immediately follows [22], and the correction with $n$=1 gives the result of a Mehra. In another limiting case $d/\delta_T << 1$, Schwinger refers to work [27] showing the impossibility of using the Lifshitz formula for the final temperature due to its non-analyticity at $d/\delta_T \to 0$. He recognizes the inaccuracy of this formula for extremely small $d$, the impossibility of a limiting transition, summarizes the series using the Poisson formula and obtains Mehra's result for low temperatures. However, he uses the formula

$$\int_0^\infty \frac{\sin(zy)dy}{\exp(y)-1} = \frac{\pi}{2}\coth(\pi z) - \frac{1}{2z},$$

which is the difference of two infinite quantities at $z = 2\pi n/t$, $t = 4\pi d/(k_B T) \to 0$, while changing the order of integration. In this case, the final result can change by any large amount. It was shown in [27] that formula (12) contains an error (see also [4, 28]). Since there was no consistent derivation of formulas in [10], Lifshitz's formulas were initially questioned (see, for example, [7]). However, the derivation of formula (4) in a number of papers, as well as experiments, have confirmed its correctness. Lifshitz tried to derive the formula by introducing proportional source dissipation $\varepsilon''(\omega)\Theta(\omega,T)$ into Maxwell's equations (formula (1.1) from [10]), following Rytov's approach [11]. However, the final formulas did not include the dissipation $\varepsilon''(\omega)$, but only the reflection coefficients and $\Theta(\omega,T)$. The conclusion based on the Rytov method requires the use of correlation relations obtained taking into account the shape of interacting bodies based on solving problems about their excitation by dipoles [11,29,30]. The introduction of thermal fluctuations as sources in Maxwell's equations is advisable in the case of high temperatures during the interaction in the far zone of bodies that slightly change the boundary conditions. These are, for example, remote carbon nanotubes with radii of the order of 1 nm. In this case, the radiation of linear current sources into the far zone can only be considered, taking into account multiple mutual radiations. By writing down the usual correlation relations for fluctuating current densities [11]

$$\left\langle j_\alpha(\mathbf{r}), j_\beta(\mathbf{r}')\right\rangle = \frac{\omega\Theta(\omega,T)\varepsilon_0\varepsilon''(\omega)}{4\pi^2}\delta_{\alpha\beta}\delta(\mathbf{r}-\mathbf{r}'),$$

$\alpha,\beta = x,y,z$, turning to the linear conductivity $\zeta$, we obtain the correlation relations $\left\langle I_n(z), I_m(z')\right\rangle = \Theta(\omega,T)\mathrm{Re}(\varsigma(\omega))\delta_{nm}\delta(z-z')$, $n,m=1,2$, expressing the fact that the currents in different sections of the tubes do not correlate. Correlations occur after multiple re-reflections, when the emitted waves return back [32,32]. At the same time, the thermal part of the force decreases more slowly than the quantum vacuum one and significantly exceeds it.

So, the result of this work is as follows. The dispersion forces in plane-layered structures are determined by the dispersion equations for electric and magnetic PP, and the space *d* between the structures can be filled with an arbitrary dielectric. Fast radiated PP are responsible for the force in the far zone, whereas slow surface ones determine the force at short distances, since they exponentially attenuate with its increase [33, 34]. At zero temperature, the force is determined by the Schram formula, which coincides with the Lifshitz formula in the case of a vacuum gap between dielectric half-spaces. Without taking into account the delay, the formula turns into the van Kampen–Schram formula. For a nonzero temperature, the force is given by the Lifshitz formula for a finite temperature plus the contour integral in the sense of the principal value, which reduces to an integral over the variable $\xi = i\omega$, which includes $\cot(\xi\delta_T/(2c))$. This integral can be reduced to an integral in a finite domain, understood in the usual sense. The Lifshitz temperature formula at room temperature is very accurate at distances between structures of more than 100 nm. For smaller distances, the contour integral needs to be taken into account.

**Financing.** The work was carried out with the financial support of the Ministry of Education and Science of the Russian Federation within the framework of the state assignment (FSRR-2026-0006).